\documentclass[conference]{IEEEtran}

\usepackage{amsmath,amssymb,amsfonts}
\usepackage{graphicx}
\usepackage{textcomp}
\usepackage{xcolor}

\usepackage{booktabs}      
\usepackage{enumitem}      
\usepackage{siunitx}       
\usepackage{xspace}
\usepackage{url}
\usepackage{stfloats}      
\usepackage{tikz}          
\usepackage[capitalize,noabbrev]{cleveref}   
\graphicspath{{figures/}{./figures/}}

\crefname{figure}{Fig.}{Figs.}
\Crefname{figure}{Fig.}{Figs.}
\crefname{table}{Table}{Tables}
\Crefname{table}{Table}{Tables}
\crefname{section}{Section}{Sections}
\Crefname{section}{Section}{Sections}

\makeatletter
\def\@IEEEsectpunct{}      
\def\paragraph{\@startsection{paragraph}{4}{\z@}%
  {1.3ex plus 0.4ex minus 0.2ex}{-0.5em}{\normalfont\normalsize\bfseries}}
\makeatother

\usepackage{makecell}
\usepackage{multirow}

\newcommand{\oursystem}{\texttt{GrayTrack}\xspace}
\IEEEoverridecommandlockouts
\begin{document}

\title{\fontsize{22}{26}\selectfont Beyond Direct Sensing: Harnessing Indirect Observations from Third-Party Sensors in Vehicle Tracking}

\author{%
  Gaofeng~Dong$^{*}$, 
  Vamsi~Eyunni$^{*}$, 
  Pragya~Sharma, 
  Kang~Yang, 
  Mani~Srivastava  \\
    Department of Electrical and Computer Engineering \\
    University of California, Los Angeles \\
    Los Angeles, CA, USA \\
  \{gfdong, veyunni, pragyasharma, kyang73, mbs\}@ucla.edu
  \vspace{-0.1in}

\thanks{$^{*}$Equal contribution.}
\thanks{Mani Srivastava holds concurrent appointments as a Professor of ECE and CS (joint) at the University of California, Los Angeles.}
\thanks{© 2026 IEEE.  Personal use of this material is permitted.  Permission from IEEE must be obtained for all other uses, in any current or future media, including reprinting/republishing this material for advertising or promotional purposes, creating new collective works, for resale or redistribution to servers or lists, or reuse of any copyrighted component of this work in other works.}
}

\maketitle

\begin{abstract}
Vehicle tracking is fundamental to applications ranging from urban mobility and public safety to security and defense. Conventional tracking relies on direct access to sensors that provide strong observations such as vehicle identity and location. In practice, however, factors such as ownership, privacy, cost, and operational constraints may limit directly accessible sensors, leaving sparse observations and long tracking gaps. Meanwhile, many additional third-party sensing assets may be present across the environment but remain inaccessible at the raw-data level, preventing their direct integration into the tracking system. In this work, we investigate whether weak, indirect observations with uncertain spatial and temporal cues can complement sparse direct sensing for vehicle tracking. Specifically, we propose \oursystem, which fuses weak anonymous events with sparse direct observations using a road-constrained particle filter. We build a CARLA–Mininet-WiFi pipeline to evaluate the system under controlled conditions, generating direct observations from accessible cameras and indirect observations from third-party cameras. Our learning-based detector achieves an F1 score of 0.989 for anonymous vehicle passages. Further, incorporating indirect third-party observations reduces trajectory RMSE by 60.1\% and catastrophic track loss from 35.8\% to 0.3\%. These results demonstrate that \oursystem can effectively exploit weak indirect observations to extend tracking capabilities.

\end{abstract}



\section{Introduction}
\label{sec:intro}

Vehicle tracking is critical across civilian and military applications, including urban mobility, public safety, and battlefield situational awareness. Such systems typically rely on sensors under the operator's control or otherwise authorized for direct access, such as cameras that provide high-quality observations of vehicle identity, location, and time. In practice, however, ownership, privacy, encryption, cost, and operational constraints often limit direct access to only a sparse subset of the available sensing infrastructure, leaving long observation gaps in which vehicle location becomes increasingly uncertain.

\begin{figure}[th]
\centering
\includegraphics[width=0.45\textwidth]{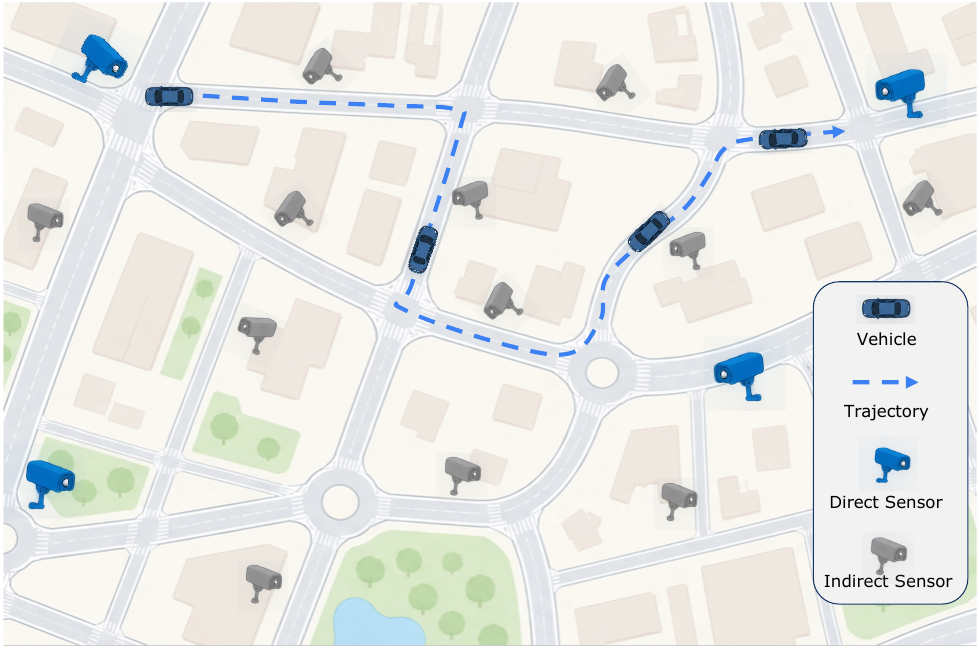}
\caption{Vehicle tracking using sparse direct observations and weak anonymous observations from gray sensing assets.}
\label{fig:scenario}
\vspace{-0.25in}
\end{figure}

Urban camera networks illustrate this gap between observation quality and coverage. As shown in Fig.~\ref{fig:scenario}, a city may deploy a limited number of public or authorized identity-capable cameras, such as automatic licence-plate readers, concentrated at selected intersections, poles, and highway ramps~\cite{monahan2026flock}. Meanwhile, the same environment may contain many more third-party sensing assets: Atlanta's Connect Atlanta program, for example, encompasses tens of thousands of cameras contributed by businesses, residents, and other organizations~\cite{scott2026atlantacameras}. Although raw observations from such assets may be unavailable due to ownership, privacy, or access restrictions~\cite{lipton2023fusus}, they may still provide weaker cues about activity in their vicinity. We refer to such externally owned or indirectly accessible sensing infrastructure as \emph{gray assets}~\cite{yetim2026tracking}.

Rather than requiring gray assets to provide vehicle identity, appearance, or precise location, we consider a weak anonymous passage event: an observation indicating only that some vehicle passed a known location at approximately a given time. The downstream tracker operates on this generic event-level interface, which could be provided by motion or proximity sensors, RF-based sensing, or deployments exposing only anonymous presence events rather than raw sensing data. This allows heterogeneous and geographically distributed sensing assets to contribute through a common interface.
In this work, we consider the common case where the third-party sensing assets are networked cameras, as illustrated in Fig.~\ref{fig:scenario}. Even when video content is encrypted, observable packet timing and traffic volume can reveal motion-related information~\cite{yetim2026tracking,rasool2025invisible}. We use these features only to infer anonymous vehicle-passage events, without circumventing encryption or accessing video content.

The challenge is that these indirect observations are inherently weak and noisy: they provide only coarse spatiotemporal evidence, cannot identify which vehicle generated an event, and may be missing, delayed, or spurious. Vehicle motion, however, is strongly constrained by physical structure. Road connectivity and plausible travel times can eliminate trajectory hypotheses that would otherwise remain feasible, suggesting that even weak local events may provide useful constraints during gaps between sparse direct observations. This leads to our central question: 
\textit{can indirect observations from third-party sensors meaningfully complement sparse direct observations for vehicle tracking?}

We address this question with \textbf{\oursystem}, a vehicle-tracking framework that combines anonymous passage events from gray assets with sparse direct observations under road-network constraints. We instantiate the event interface using encrypted third-party camera traffic and evaluate \oursystem with a CARLA--Mininet-WiFi pipeline connecting measured passage-detection errors to controlled experiments.

\paragraph{Contributions.}
Our key contributions are summarized as follows.
First, we introduce an event-level abstraction for gray-asset sensing and show that encrypted network traffic from third-party cameras can be transformed into anonymous vehicle-passage events using learning-based packet grouping and vehicle-visibility detection.
Second, we develop a road-constrained particle-filtering framework that fuses these weak, identity-free events with sparse direct observations. In our evaluation, it reduces trajectory RMSE by 60.1\% and catastrophic track loss from 35.8\% to 0.3\%, while outperforming unconstrained Kalman filtering by 27.6\%.
Third, we build a CARLA--Mininet-WiFi pipeline and conduct controlled experiments with held-out sensing sequences and simulated region-transit trajectories, systematically evaluating passage detection, sensing noise, road constraints, blind gaps, and multi-vehicle ambiguity.
\textbf{Project webpage:} \url{https://nesl.github.io/GrayTrack/}

\section{Related Work}
\label{sec:related}

\paragraph{Encrypted traffic as a sensing channel.}
Prior work has inferred device activity and household behaviour from IoT traffic. 
More recent work has shown that encrypted wireless traffic can reveal motion-related information and support physical-world inference~\cite{yetim2026tracking, rasool2025invisible}. These studies establish encrypted traffic as a viable side channel for sensing beyond network-layer activity.
Our goal differs in the information we extract and the system role assigned to the side channel. Rather than using encrypted traffic as the primary source for detailed activity or geospatial inference, we deliberately reduce each third-party sensor to a minimal anonymous passage event: a vehicle was observed near a known camera at approximately a given time. This abstraction avoids dependence on vehicle identity, appearance, or detailed scene reconstruction and is well suited to geographically distributed third-party sensors across a road network or large area. In contrast to prior settings that jointly observe a localized area, such as an intersection, using multiple cameras~\cite{yetim2026tracking}, our sensors may be spatially separated and individually provide only weak local evidence. The side channel is therefore not the final inference target; instead, we ask whether these indirect events can complement sparse identity-bearing direct observations when fused with physical constraints from the road network.

\paragraph{Multi-camera vehicle tracking and data association.}
Traditional multi-camera tracking associates observations across sensors using spatial-temporal consistency, multiple-hypothesis reasoning~\cite{reid1979mht, yao2022city}, or visual features such as vehicle appearance and re-identification~\cite{liu2016veri, nguyen2023multi}. Our third-party observations provide none of these appearance or identity cues: an event reports only that a vehicle passed a particular sensing location at approximately a given time. Consequently, association must rely on the vehicle's prior state, timing, and physical feasibility rather than visual re-identification. This makes our setting complementary to conventional multi-camera tracking: instead of extracting richer features from accessible video, we investigate how much tracking information can be recovered when most intermediate sensors provide only indirect observations.

\section{System and Problem Formulation}
\label{sec:setting}
We consider vehicle tracking using two classes of observations that differ in accessibility and information content: sparse \emph{direct observations} from accessible sensors and weaker \emph{indirect observations} from third-party sensors.

\textbf{Direct sensing.}
Directly accessible sensors provide vehicle-specific direct observations $(t,x,y,\mathrm{id})$, containing a timestamp, location, and vehicle identity. These observations are informative but sparse because the observer has direct access to only a subset of the sensing infrastructure. 

\textbf{Indirect sensing.}
Third-party sensors are not directly accessible to the tracker. In this work, we consider the common case in which these sensors are networked cameras transmitting encrypted video over wireless links. Following prior work~\cite{yetim2026tracking,rasool2025invisible}, we assume that the observer deploys wireless traffic monitors that passively observe transmissions from nearby third-party cameras. A monitor may observe multiple cameras within wireless range without requiring line of sight. Observed traffic is associated with a known camera $s$, whose location $(x_s,y_s)$ is obtained from deployment metadata or registration, or inferred through prior calibration. The monitor retains only packet timing and size, without accessing the underlying video content.
From this traffic, \oursystem infers an anonymous passage event $(\hat{t},s)$ indicating that a vehicle passed near camera $s$ around time $\hat{t}$. Each event provides only coarse spatiotemporal evidence and may be missed, spurious, or temporally perturbed. The downstream tracker operates only on this event-level representation. In deployments where third-party sensor owners voluntarily expose anonymous presence or passage events, \oursystem could consume them directly without wireless traffic monitoring, although we do not rely on such cooperation in this work.

Given sparse direct observations and indirect passage events, our goal is to continuously estimate vehicle position under road-network constraints.

\paragraph{Tracking problem.}
Let the road network be a directed graph $G=(V,E)$, where $V$ denotes road junctions and $E$ directed lane segments. At time $t$, the vehicle state is
\[
    \mathbf{x}_t = (\mathbf{p}_t, v_t),
    \qquad
    \mathbf{p}_t=(x_t,y_t),
\]
where $\mathbf{p}_t$ is the vehicle position and $v_t$ its speed. Vehicle motion
is constrained by the road graph $\mathbf{p}_t \in G$
such that feasible trajectories follow connected road segments. The observer receives two asynchronous observation streams. Direct sensors provide sparse direct observations
\[
    a_i = (t_i,\mathbf{p}_i,\mathrm{id}),
    \qquad a_i \in \mathcal{A},
\]
while third-party sensors provide indirect passage events
\[
    e_j = (\hat{t}_j,s_j),
    \qquad e_j \in \mathcal{E},
\]
where $s_j$ has known location $\mathbf{c}_{s_j}=(x_{s_j},y_{s_j})$.
An indirect event $e_j$ provides only uncertain evidence that a vehicle was near sensor $s_j$ around time $\hat{t}_j$.

Given the road graph $G$, direct-observation stream $\mathcal{A}{1:t}$, and indirect-observation stream $\mathcal{E}{1:t}$ up to time $t$, the tracking objective is to estimate
\[
    p\!\left(
        \mathbf{x}_t
        \mid
        \mathcal{A}_{1:t},
        \mathcal{E}_{1:t},
        G
    \right),
\]
and reconstruct the continuous trajectory $\hat{\mathbf{X}}_{0:T}=\{\hat{\mathbf{x}}_t\}_{t=0}^{T}$.

\section{Method}
\label{sec:method}

\begin{figure}[ht]
\centering
\includegraphics[width=0.4\textwidth]{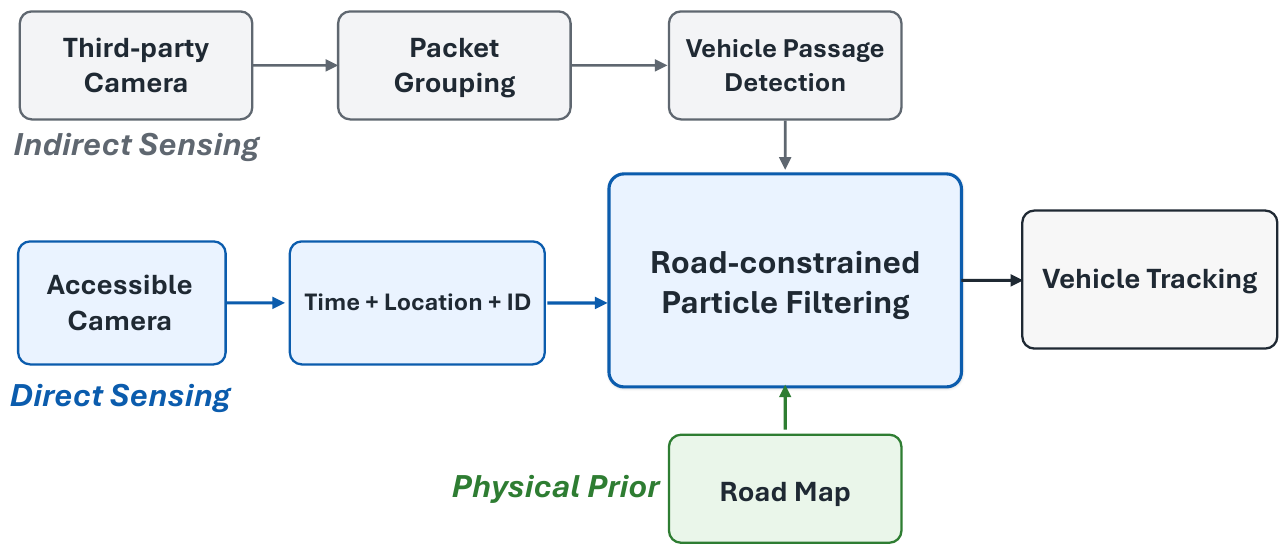}
\caption{\oursystem system overview: a road-constrained particle filter fuses indirect observations with sparse direct observations for vehicle tracking.}
\label{fig:system}
\vspace{-0.15in}
\end{figure}

Fig.~\ref{fig:system} summarizes \oursystem. Encrypted traffic from third-party cameras is processed into indirect vehicle-passage observations, which are fused with sparse direct observations under road-network constraints to estimate continuous vehicle trajectories. Direct observations from accessible cameras can be obtained using established vision methods, such as vehicle detection and license-plate recognition. We therefore focus on the indirect sensing path.

\subsection{Passage Detection from Encrypted Camera Traffic}
\label{sec:method-detector}

For each third-party camera, we passively collect network traffic and retain only packet arrival times and sizes, without accessing the payloads or video content, to infer the occurrence and timing of vehicle pass-by events. Video codecs use inter-frame prediction, causing scene changes to alter encoded frame sizes that remain observable in encrypted traffic~\cite{yetim2026tracking,rasool2025invisible}. Because frames may span multiple packets, we first recover frame boundaries from packet metadata and then detect vehicle visibility from the resulting frame-size sequence~\cite{yetim2026tracking}.

\paragraph{Stage 1: Packet grouping.}
For camera $s$, let
$\mathcal{P}_s=\{(\tau_i,b_i)\}_{i=1}^{N_s}$,
where $\tau_i$ and $b_i$ denote the arrival time and size of network packet $i$.
A learned packet-grouping model predicts whether each network packet marks the end of an encoded video frame:
\[
    \hat{m}_i =
    f_{\theta}\!\left(
        \{(\tau_k,b_k)\}_{k\in\mathcal{W}_i}
    \right),
\]
where $\mathcal{W}_i$ is a temporal packet window and $\hat{m}_i$ is the predicted frame-boundary indicator. These boundaries partition the packets into inferred frames $\mathcal{G}_1,\ldots,\mathcal{G}_M$. We recover the size of frame $j$ as
\[
    B_j=\sum_{i\in\mathcal{G}_j} b_i,
\]
yielding the frame-level sequence
$\mathcal{F}_s=\{(\hat{\tau}_j,B_j)\}_{j=1}^{M}$.

\paragraph{Stage 2: Vehicle passage detection.}
For each inferred frame in $\mathcal{F}_s$, we construct a feature vector $\mathbf{z}_j$ from its frame size $B_j$ and local temporal context, including log-transformed frame size, exponentially weighted residual and ratio features, and frame time span. These features reduce sensitivity to absolute bitrate and transient network variation and instead emphasize relative changes in encoded frame complexity, reducing reliance on camera-specific absolute bitrate characteristics.
A second learned model estimates vehicle visibility over a temporal window:
\[
    \hat{y}_j =
    g_{\phi}\!\left(
        \{\mathbf{z}_k\}_{k\in\mathcal{V}_j}
    \right),
\]
where $\mathcal{V}_j$ is the temporal window around frame $j$, and $\hat{y}_j$ is the predicted probability that a vehicle is visible in frame $j$. Thus, although inference uses temporal context, visibility is predicted frame by frame. Consecutive positive predictions are merged into visibility intervals, each producing a passage event $e_q=(\hat{t}_q,s)$. The downstream tracker uses only these event-level observations, rather than the underlying packet or frame representations.
For event-level evaluation, predicted and ground-truth passages are matched one-to-one using their interval midpoints within a matching tolerance.

\subsection{Indirect-Observation Model}
\label{sec:method-obsmodel}

We evaluate passage detection separately and use its measured event-level data to parameterize indirect observations in the tracking experiments. Starting from oracle events $\mathcal{E}^{*}$ generated from ground-truth trajectories and sensor geometry, we construct an observed stream $\tilde{\mathcal{E}}$ that models missed detections, false positives, and timing uncertainty. 
Each true event is retained with probability $1-p_{\mathrm{miss}}$, false events are introduced according to a per-passage rate $\lambda_{\mathrm{false}}$, and retained event timestamps are perturbed as
\[
    \hat{t}_j=t_j^{*}+\delta_j,
    \qquad
    \delta_j\sim\mathcal{N}(\mu_t,\sigma_t^2).
\]
The nominal parameters are derived from the measured passage-detector performance in \cref{sec:rq1}, integrating the sensing and tracking experiments while independently varying each source of error.

\subsection{Road-Constrained Particle Filter}
\label{sec:method-pf}

We fuse direct and indirect observations using a bootstrap particle filter~\cite{gordon1993novel}. Each particle represents a vehicle hypothesis directly on the road graph $G=(V,E)$, with state
\[
    \mathbf{x}_t^{(i)}
    =
    \bigl(e_t^{(i)},\ell_t^{(i)},v_t^{(i)}\bigr),
\]
where $e_t^{(i)}\in E$ is the current lane segment, $\ell_t^{(i)}$ is the position along the segment, and $v_t^{(i)}$ is the vehicle speed. Between observations, particles propagate along the road with stochastic velocity perturbations; at junctions, they transition among feasible outgoing segments, maintaining multiple route hypotheses while remaining road-constrained.

Direct observations concentrate particles around the observed vehicle location. An indirect event $e_j=(\hat{t}_j,s_j)$ instead provides weaker evidence that the vehicle passed near sensor $s_j$ at known location $\mathbf{c}_{s_j}$. We model its likelihood for particle $i$ as
\[
    p(e_j\mid \mathbf{p}_{\hat{t}_j}^{(i)})
    \propto
    \exp\!\left(
        -\frac{
        \|\mathbf{p}_{\hat{t}_j}^{(i)}-\mathbf{c}_{s_j}\|_2^2
        }{2\sigma_{\mathrm{loc}}^2}
    \right),
\]
where $\mathbf{p}_{\hat{t}_j}^{(i)}$ is the particle position and $\sigma_{\mathrm{loc}}$ captures spatial uncertainty. Thus, an indirect event reweights rather than relocates particles, favoring road-consistent hypotheses that pass near the reporting sensor at a compatible time. We then normalize and resample particle weights following the standard bootstrap particle-filter procedure, with the weighted particle distribution providing the estimated vehicle position.

\section{Evaluation}
\label{sec:results}

\subsection{Experimental Setup}
\label{sec:setup}

\paragraph{Implementation.}
Passage detection is evaluated on encrypted wireless camera traffic collected using our CARLA–Mininet-WiFi pipeline, where CARLA~\cite{dosovitskiy2017carla} generates vehicle motion and camera streams and Mininet-WiFi~\cite{fontes2015mininetwifi} emulates encrypted wireless transmission. We collect 2,607 camera traces across 237 runs and 11 cameras. Event-level passage detection is evaluated on 242 held-out camera sequences. Then we conduct controlled downstream tracking experiments on \num{120} simulated region-transit trajectories over the CARLA Town05 road network, totaling \SI{35.9}{\kilo\meter} of vehicle travel. The region contains \num{12} sparse direct-sensing perimeter sensors and \num{15} indirect interior cameras. Indirect observations are parameterized by the measured event-level detector errors. Unless otherwise stated, the main tracking experiments consider a single vehicle at a time; concurrent vehicles are evaluated separately in \cref{sec:multivehicle}.
The packet-grouping model, PacketSegformer, uses a four-layer Transformer encoder ($d_{\mathrm{model}}=16$) over 250-packet windows, using packet timing and size together with derived temporal features to predict frame boundaries. The vehicle-visibility detector, CarVisibleLSTM, uses a two-layer unidirectional LSTM with 64 hidden units over 512-frame windows and predicts frame-level vehicle visibility from frame-size and temporal features. Passage events are obtained by dropping short detections and fusing nearby intervals within a specified merge gap. For downstream tracking, Road-PF uses \num{2000} particles.

\paragraph{Evaluation conditions.}
We evaluate three sensing conditions: \textbf{A}, sparse direct observations only; \textbf{B}, sparse direct observations plus oracle indirect passage events, representing an \textit{upper bound} on the benefit of indirect sensing; and \textbf{C}, sparse direct observations plus indirect passage events corrupted according to the measured event-level detector errors, representing a realistic noisy scenario.

\paragraph{Baselines.} 
For passage detection, we compare PacketSegformer with fixed \SI{50}{\milli\second} packet binning at the known \SI{20}{\hertz} frame rate, and a rule-based detector with a hand-tuned median\,+\,$6\times$MAD frame-size threshold. For tracking, we compare Road-PF against dead reckoning and an unconstrained Kalman filter receiving the same direct and indirect observations. Dead reckoning propagates the latest observed state using its estimated velocity. The Kalman filter uses a constant-velocity model in 2D and treats passage events as uncertain position observations centered at the reporting sensor. Kalman and Road-PF use the same observation uncertainty, isolating the effect of road-network constraints.

\paragraph{Metrics.}
For passage detection, we report precision, recall, and F1 at the frame-boundary and passage-event levels. For tracking, we evaluate continuous trajectory reconstruction using per-trajectory position RMSE over the entire transit. We additionally report the \emph{catastrophic-failure rate}, defined as the fraction of trajectories with RMSE exceeding \SI{100}{\meter}, indicating that the estimated trajectory is no longer in the correct neighbourhood.

\subsection{Passage Detection from Encrypted Camera Traffic}
\label{sec:rq1}

We first evaluate whether encrypted packet metadata can be reliably transformed into the indirect passage observations required by the downstream tracker. We evaluate the two stages separately: network packet grouping measures recovery of encoded video frame boundaries, while vehicle-visibility detection evaluates whether the resulting frame-size dynamics accurately predict vehicle passages.

\begin{table}[t]
\centering
\caption{Performance of frame-boundary recovery and passage-event detection.}
\label{tab:detection}
\footnotesize
\setlength{\tabcolsep}{3.5pt}
\begin{tabular}{llccc}
\toprule
Task & Method & Precision & Recall & F1 \\
\midrule
\multirow{2}{*}{\makecell{Frame-boundary\\detection}}
    & Fixed-period     & 0.926 & 0.838 & 0.880 \\
    & PacketSegformer & 0.997 & 0.999 & 0.998 \\
\midrule
\multirow{2}{*}{\makecell{Passage-event\\detection}}
    & Rule-based pipeline & 0.885 & 0.900 & 0.893 \\
    & Learning-based pipeline   & 0.978 & 1.000 & 0.989 \\
\bottomrule
\end{tabular}
\vspace{-0.2in}
\end{table}

PacketSegformer achieves an F1 score of 0.998 for frame-boundary detection, compared with 0.880 for the fixed-period baseline (\cref{tab:detection}). The improvement is driven primarily by substantially higher recall. 
More importantly for downstream detection, the recovered boundaries yield a normalized frame-size MAE of only 0.0025, compared with 0.8192 for the fixed-period baseline.
The learning-based pipeline improves passage-event detection over the rule-based baseline, increasing F1 from 0.893 to 0.989. The rule-based approach heavily relies on hand-tuned, sequence-specific calibration, significantly limiting its scalability across heterogeneous deployments. We therefore use the learning-based pipeline for downstream experiments. Its event-level errors correspond to a missed-event rate of 0.0, a false-event rate of 0.0225 per true passage, and passage-time errors with mean bias 0.022 s and standard deviation 0.109 s; these measurements parameterize the indirect-observation model used in downstream tracking.

\subsection{Tracking with Indirect Observations}
\label{sec:rq2}

We first ask whether indirect third-party observations provide meaningful tracking value beyond sparse direct sensing.
Between two direct observations, a vehicle remains unobserved and its trajectory uncertainty grows over time.
Anonymous passage events provide intermediate evidence about where the vehicle may have traveled, while the road network further restricts the set of physically feasible trajectories.

With direct observations alone, the road-constrained filter achieves a mean trajectory RMSE of \SI{92.3}{\meter}.
Incorporating indirect passage events at the measured detector error rates reduces RMSE to \SI{36.8}{\meter}, a \SI{60.1}{\percent} reduction, close to the \SI{35.8}{\meter} RMSE achieved with oracle passage events.
Thus, even noisy indirect observations recover a substantial fraction of the tracking benefit available from perfect intermediate sensing.
The improvement is even more pronounced for catastrophic failures.
The fraction of trajectories with RMSE above \SI{100}{\meter} decreases from 35.8\% with direct anchors alone to 0.3\% with noisy indirect observations, nearly matching the 0\% achieved with oracle events.
These results show that indirect third-party observations can substantially reduce both average trajectory error and severe track loss between sparse direct observations.

\subsection{Robustness to Noisy Indirect Observations}
\label{sec:rq3}

\begin{figure*}[t]
\centering
\includegraphics[width=0.8\textwidth]{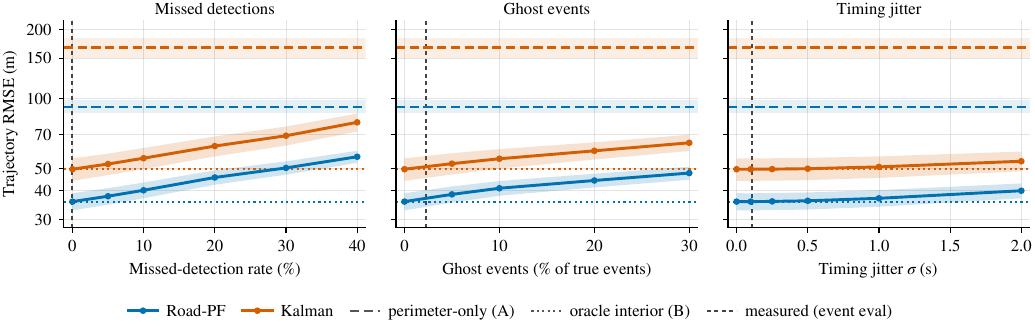}
\vspace{-0.1in}
\caption{Robustness to indirect-observation errors. Each error source is varied independently while the others are held at their measured nominal values. Horizontal references show direct sensing only (A) and oracle indirect observations (B), and vertical markers indicate the operating points from \cref{sec:rq1}.}
\label{fig:robustness}
\vspace{-0.2in}
\end{figure*}

We next evaluate how tracking performance degrades as indirect observations become less reliable. In our camera-based realization, errors can arise from passive wireless traffic monitoring, network variability, and event inference, resulting in missed passages, false events, or uncertain timestamps. We independently vary the missed-detection rate, false-event rate, and timing jitter while holding the other error sources at their measured nominal values, isolating the effect of each source.
As shown in \cref{fig:robustness}, tracking error increases gradually with each error source, and Road-PF remains substantially better than direct sensing alone across all evaluated noise levels. 
Even at severe settings tested, indirect observations continue to provide useful intermediate constraints on the vehicle trajectory.

The system is not equally sensitive to all error sources. Increasing missed detections and ghost events produces the largest degradation, while moderate timing uncertainty has a smaller effect on trajectory accuracy.
Road-PF also consistently outperforms the unconstrained Kalman filter throughout the sweeps. Overall, these results show that the tracking benefit persists under substantial indirect-observation errors and does not depend on unrealistically accurate sensing.

\subsection{Estimator Comparison and Blind Gaps}
\label{sec:rq4}

\begin{table}[t]
\centering
\caption{Tracking performance under the three sensing conditions. Catastrophic failure denotes trajectory RMSE $> \SI{100}{\meter}$.}
\label{tab:estimators}
\footnotesize
\setlength{\tabcolsep}{5pt}
\begin{tabular}{lcc}
\toprule
Estimator & RMSE (\si{\meter}) & Failure rate \\
\midrule
\multicolumn{3}{l}{\emph{A. Direct observations only}} \\
Dead reckoning, Kalman filter & 166.8 & 74.2\% \\
\textbf{Road-PF} & \textbf{92.3} & \textbf{35.8\%} \\
\midrule
\multicolumn{3}{l}{\emph{B. Direct observations + oracle indirect events (upper bound)}} \\
Dead reckoning & 48.0 & 6.7\% \\
Kalman filter & 49.6 & 8.3\% \\
\textbf{Road-PF} & \textbf{35.8} & \textbf{0.0\%} \\
\midrule
\multicolumn{3}{l}{\emph{C. Direct observations + noisy indirect events}} \\
Dead reckoning & 50.5 & 8.3\% \\
Kalman filter & 50.8 & 8.8\% \\
\textbf{Road-PF} & \textbf{36.8} & \textbf{0.3\%} \\
\bottomrule
\end{tabular}
\vspace{-0.2in}
\end{table}

We next evaluate how much road-network constraints contribute beyond the indirect observations. 
Under the calibrated noisy-observation setting, Road-PF achieves \SI{36.8}{\meter} RMSE, a 27.6\% reduction relative to Kalman filtering, with the lowest catastrophic-failure rate across all estimators.
The road-constrained filter achieves the lowest error across all three sensing conditions, showing that road topology provides useful constraints both with and without indirect observations.

To understand when this advantage is most pronounced, we test two possible explanations. One possibility is that road constraints primarily help resolve ambiguity at junctions; however, we find no clear relationship between branch density and the particle filter’s advantage ($p=0.90$). 
In contrast, grouping trajectories by their longest gap between direct observations, the Road-PF advantage over Kalman filtering increases from \SI{5.4}{\meter} on short-gap routes to \SI{26.8}{\meter} on long-gap routes ($p<0.001$).
These results indicate that road constraints are particularly valuable during long predict-only intervals, when unconstrained propagation accumulates drift while road-constrained particles remain confined to physically feasible trajectories.

\subsection{End-to-end Validation}
Our primary evaluation deliberately separates passage detection from downstream tracking for controlled analysis. End-to-end replay validates this abstraction: Road-PF RMSE drops from 78.0 m with direct observations only to 44.3 m with detected events, closely matching 44.1 m with oracle events. This result is consistent with our controlled evaluation and confirms that the measured detector performance transfers effectively to downstream tracking.

\subsection{Multi-Vehicle Case Study}
\label{sec:multivehicle}

We finally evaluate \num{15} scenarios each with one, two, and three concurrently active vehicles, with overlapping transit intervals to induce data-association ambiguity among passage events. 
Direct observations carry vehicle identity and are applied directly to the corresponding track, whereas indirect events are assigned using gated Hungarian matching based on the distance between each track's predicted mean position and the reporting sensor. Once assigned, Road-PF updates the corresponding particles using the likelihood in Section~\ref{sec:method-pf}; the Kalman baseline uses greedy likelihood-based association followed by a standard measurement update.

For the road-constrained filter, event-to-track assignment accuracy decreases from 1.00 with one vehicle to 0.87 with two and 0.63 with three, showing increasing ambiguity with vehicle density. 
This ambiguity becomes more challenging as the number of vehicles increases and the closest inter-vehicle separation decreases. 
At three vehicles, Road-PF achieves higher assignment accuracy (0.63 versus 0.58) and fewer identity switches (2.5 versus 3.1) than the Kalman filter. Road-PF also significantly reduces trajectory RMSE (69.2 versus 82.6 m) and track-loss rate (0.15 versus 0.33). These results highlight anonymous event-to-vehicle association as an important remaining challenge in multi-vehicle settings, while road constraints continue to improve overall tracking accuracy and robustness. Our multi-vehicle study focuses on low-to-moderate traffic regimes, which are relevant to localized monitoring scenarios such as perimeter crossings, sparse road networks, rural areas, and mission-critical operating zones. Even in these ambiguous settings, indirect observations provide intermediate constraints during gaps between sparse direct observations, where tracking would otherwise rely largely on motion and road-network priors alone.

\section{Discussion and Future Directions}
\label{sec:limits}
This work uses encrypted camera traffic to obtain indirect observations. More generally, the downstream tracker operates on a generic event-level interface that can support other sensing modalities, such as motion or proximity sensors, as well as privacy-preserving deployments exposing only anonymous presence or passage events.
In addition, the learning-based models are trained offline using frame-boundary and vehicle-visibility labels. At deployment time, inference requires only packet timing and size and does not require access to the underlying video content. Improving generalization to unseen cameras, thereby reducing or eliminating the need for labeled calibration data, remains a direction for future work, including training on data from more diverse codecs and camera viewpoints and developing more generalizable models.
For multi-vehicle settings, future work can explore scalable probabilistic or multi-hypothesis data association for anonymous events.

\section{Conclusion}
\label{sec:conclusion}

We presented \oursystem, a vehicle-tracking framework that combines sparse direct observations with third-party indirect observations from gray assets under road-network constraints. Using encrypted camera traffic as one realization of this event-level interface, \oursystem reduces trajectory RMSE by 60.1\% and catastrophic track loss from 35.8\% to 0.3\%.
More broadly, our results demonstrate the value of indirect sensing for extending tracking beyond the coverage of directly accessible sensors.

\section{Acknowledgment}
The research reported in this paper was sponsored in part by: the DEVCOM Army Research Laboratory under award \#W911NF1720196; the National Science Foundation under awards \#CNS-2211301 and ECCS-2525614; and Sandia National Laboratories under award \#2169310. The views and conclusions contained in this document are those of the authors and should not be interpreted as representing the official policies, either expressed or implied, of the funding agencies.

\bibliographystyle{IEEEtran}
\bibliography{./tex/references}

\end{document}